\documentclass[aps,prl,twocolumn,showpacs]{revtex4}
\usepackage{graphicx}
\usepackage{bm}

\usepackage{epstopdf}

\begin{document} 
\title{ Edge spin accumulation:  spin Hall effect without bulk spin current}  
\author{ E.B. Sonin}

\affiliation{ Racah Institute of Physics, Hebrew University of
Jerusalem, Jerusalem 91904, Israel} 

\date{\today} 

\begin{abstract}
Spin accumulation in a 2D electron gas with Rashba spin-orbit interaction  subject to an electric field can take place without bulk spin currents (edge spin Hall effect).  This is demonstrated for the collisional regime using the non-equilibrium distribution function determined from the standard Boltzmann equation. Spin accumulation originates from interference of incident and reflected electron waves  at the sample boundary. 
\end{abstract} 

\maketitle

Nowadays the spin Hall effect attracts a lot of attention of theorists and experimentalists because of its importance for spintronics \cite{Engel,Dy08}. Originally the spin Hall effect was defined as spin accumulation at sample edges  caused by a bulk spin current transverse to an applied electric field \cite{DP}. 
However, the very concept of spin current was a matter of debates (see Ref.~ \onlinecite{spin} for a review). There is no conservation for the total spin, and its connection with spin accumulation is not straightforward. 
Spin currents exist even in the equilibrium, though they do not lead to spin accumulation at sample edges \cite{R} but produce an edge spin torque, which can be measured mechanically \cite{ES}.  Moreover,  spin accumulation at sample edges is possible even without bulk spin current. It was demonstrated for the ballistic spin Hall effect \cite{Nik,Usaj,Zu}, when the electron mean-free path exceeds the sample sizes. Here  the natural definition of the spin current as the averaged product of the spin and group velocity is assumed.

In the ballistic regime the electric field is absent in the bulk. This makes the case rather unique  impeding possible generalizations. The present Letter demonstrates that edge spin accumulation without bulk spin currents takes place also in the standard collisional regime when the electron mean-free path is much shorter than the sizes of the sample. 
The experimental evidences of the spin Hall effect reported in the literature \cite{Kato,Wund,Nomura}, were based on measurement of the spin accumulated on the sample edges. Meanwhile, the spin accumulation is not really a probe of the bulk spin current: the former can be absent in the presence of the bulk current and can appear in the absence of the latter. This should be taken into account interpreting experiments on the spin Hall effect. 

A straightforward method to find a bulk spin current  is a solution of the Boltzmann equation. However, the standard Boltzmann equation for the scalar electron distribution function among the eigenstates of the Hamiltonian without disorder,  does not work in the case of spin normal to the sample plane. Indeed, all eigenstates of the Rashba Hamiltonian [Eq.~(\ref{Ham}) below] do not have neither the $z$ spin component nor its current. Introducing the scalar distribution function among this states one neglect any correlations between them, which makes appearance of the $z$ spin component or its current impossible. Therefore they used the quantum Boltzmann equation, in which the distribution function was a matrix 2 $\times$ 2 in spin indices  \cite{DK,Shyt}. The $z$ spin current requires finite non-diagonal terms. On the other hand, after some debates it is now generally believed that if the Rashba spin-orbit interaction is linear in the electron momentum, the bulk current of the spin  $z$ component vanishes \cite{Engel}. Since the present Letter addresses exactly this case,  non-diagonal terms of the density matrix may be neglected and one can use the standard Boltzmann equation for a scalar distribution function. 

The Letter considers  a 2D electron gas confined to a potential well with infinitely high walls.  This lead to the simplest boundary condition at the sample edges: the electron wave function must vanish.  The 2D electron gas with Rashba spin-obit interaction is described by the  single-electron Hamiltonian
\begin{eqnarray}
H ={\hbar^2\over 2 m}\left\{\vec \nabla \mathbf{\Psi}^\dagger \vec \nabla \mathbf{\Psi} 
 +i\alpha (\mathbf{\Psi}^\dagger  [\hat{\vec \sigma} \times \hat z]_i \vec\nabla_i\mathbf{\Psi} 
\right. \nonumber \\ \left.
 -\vec \nabla_i \mathbf{\Psi}^\dagger[\hat{\vec \sigma} \times \hat z]_i\mathbf{\Psi} )\right\} ,
     \label{Ham}         \end{eqnarray}
where  $\mathbf{\Psi} =\left( \begin{array}{c} \psi_\uparrow \\ \psi_\downarrow\end{array}\right)$
is a two-component spinor, $\hat{\vec \sigma}$ is the vector of Pauli matrices.
The plane-wave solutions of the Schr\"odinger equation are
\begin{eqnarray}
{1\over \sqrt{2}}\left( \begin{array}{c} 1 \\ \pm ie^{i\varphi} \end{array}\right)e^{i\vec k  \vec r},         
          \end{eqnarray} 
where $\varphi$ is the angle between the wave vector $\vec k$ and the axis $x$ ($k_x=k\cos \varphi$, $k_y=k\sin \varphi$), and the upper (lower) sign corresponds to the upper (lower) branch of the spectrum (band) with the energies   
 \begin{eqnarray}
\epsilon={\hbar^2(k_0^2-\alpha ^2)\over 2 m}  ={\hbar^2\over m}\left({k^2\over 2}\pm \alpha k\right).
           \label{ener}  \end{eqnarray}   
The energy is parameterized by the wave number  $k_0$, which is connected with absolute values of wave vectors in two bands as $k=|k_0 \mp \alpha|$.  

\begin{figure}
\begin{center}
   \leavevmode
  \includegraphics[width=0.9\linewidth]{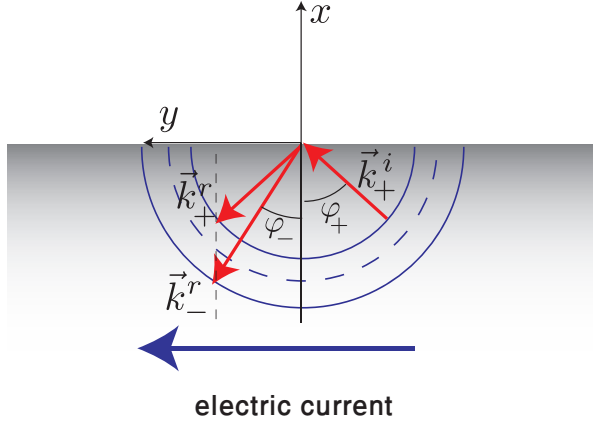}
 \caption{(Color online) Spin-dependent reflection of electrons from an ideal impenetrable wall. The electron from the upper band ($\vec k_+^i$) is reflected either as  an electron from the same band ($\vec k_+^r$), or as  an electron from the lower band ($\vec k_-^r$).}
 \label{fig1}
 \end{center}
\end{figure}

\begin{figure}
\begin{center}
   \leavevmode
  \includegraphics[width=0.95 \linewidth]{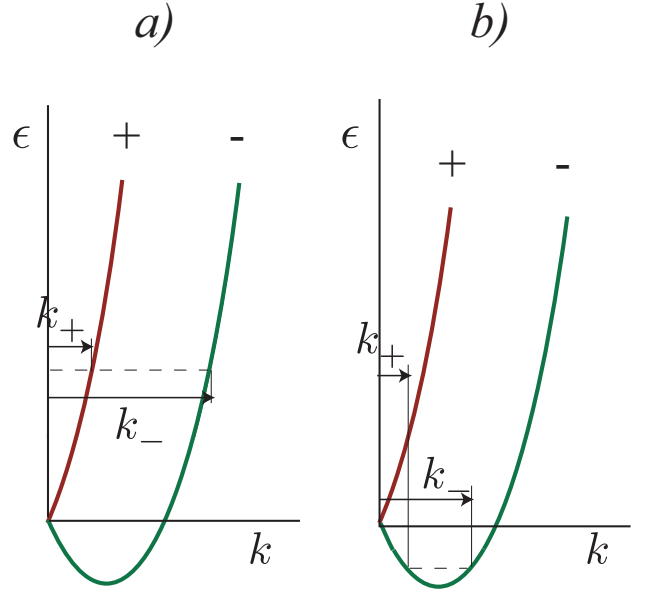}
 \caption{  The energies and the wave vectors of the waves interfering at the sample edge. a) The high-energy case $k_0 >\alpha$, the interference between electrons from the different bands, $k_\pm= k_0\mp \alpha$. b) The low-energy case $k_0 <\alpha$, the interference between electrons from the low band on two sides of the energy minimum, $k_\pm= \alpha\mp k_0$.}
 \label{fig2}
 \end{center}
\end{figure}

We assume that the 2D electron gas occupies the semispace $x<0$ (Fig.~\ref{fig1}). A superposition of plane waves, which satisfies the boundary condition ${\mathbf \Psi}(0)=0$, contains 
one incident wave coming from $x =-\infty$, and two reflected waves. For high-energy  electrons with $k_0 >\alpha $  [Fig.~\ref{fig2}(a)] and the incident electron in the upper band, the superposition is
\begin{eqnarray}
{\mathbf \Psi}={e^{ik_y y}\over \sqrt{2}}\left[\left( \begin{array}{c} 1\\ - ie^{i\varphi_+} \end{array}\right)e^{ik_{+ x} x} 
\right. \nonumber \\ \left.
+r_1\left( \begin{array}{c} 1\\  ie^{-i\varphi_+} \end{array}\right)e^{-ik_{+ x} x}
+r_2\left( \begin{array}{c} 1\\ - ie^{-i\varphi_-} \end{array}\right) e^{-ik_{- x} x} \right],  
     \label{x<}         \end{eqnarray} 
where   $ \varphi_\pm=\arctan (k_y/k_{\pm x}) $,   and  $k_{\pm x} =\sqrt{(k_0 \mp \alpha)^2-k_y^2}$ are the $x$ components of the wave vectors corresponding to states of the same energy in the upper (+) and the lower (-) band.  The reflection coefficients  for reflection to the same ($r_1$) and to the other ($r_2$) band (Fig.~\ref{fig1}) are
\begin{eqnarray}
r_1={e^{i(\varphi_+ +\varphi_-)} -1 \over e^{i(\varphi_- -\varphi_+)}+1},~~r_2=-{2i e^{i\varphi_- }\cos \varphi_+\over  e^{i(\varphi_- -\varphi_+)}+1}~.
             \end{eqnarray} 
The relation between the angles $\varphi_+$ and $\varphi_-$  is determined from the condition that scattering does not change the component $k_y=(k_0-\alpha) \sin \varphi_+ =(k_0+\alpha) \sin \varphi_- $.

We look for the density $s_z=(\hbar/2) {\mathbf \Psi}^\dagger \hat \sigma_z {\mathbf \Psi}$ of the $z$ spin component. In the plane waves  $s_z$ vanishes. But near the boundary because of the interference between the waves in the superposition a finite oscillating $s_z$ is possible (Friedel-type oscillation) and is given by
 \begin{eqnarray}
s_{+z}(\vec k) ={\hbar \over 4}\left\{r_1(e^{-2i\varphi_+}
+1)e^{-2ik_1x} 
+r_2[e^{-i(\varphi_+ +\varphi_-)}
\right. \nonumber \\ \left.
+1]e^{-i(k_{+ x}+k_{- x})x}
+r_1^* r_2[e^{i(\varphi_+ -\varphi_-)}+1]e^{i(k_{+ x}-k_{- x}) x}\right\}
\nonumber \\
+\mbox{c.c.} ={\hbar  (\sin\varphi_+ + \sin\varphi_-)\cos \varphi_+\over 1+\cos (\varphi_+-\varphi_-)}[\sin (2k_{+ x}x)
\nonumber \\
-\sin (k_{+ x}x+k_{- x}x)-\sin (k_{+ x}x-k_{- x}x)].~~
             \end{eqnarray}   
Exchanging $+$ and $-$ one obtains the spin density $s_{-z}(\vec k)$ for the incident electron from the lower band:
 
Similarly one can consider  the low-energy  case $k_0 <\alpha $ when the $x$ components of the wave vectors   $k_{\pm x} =\sqrt{(\alpha \mp k_0)^2-k_y^2}$ belong to the two states of the lower band on the left and on the right of the band energy minimum respectively [Fig.~\ref{fig2}(b)].  On the left from the minimum  the direction of the group velocity $\vec v=\partial \epsilon /\hbar \partial \vec k$ is opposite to that of the wave vector $\vec k$.  Since  the charge transport is determined by the group velocity but not by $\vec k$, the  wave vector for the incident electron from the left side  is directed from the edge but not to the edge.  

The expressions given above are valid only if $k_y < k_+$, or $\sin \varphi_- <| k_0-\alpha|/(k_0+\alpha)$.  At $k_-> k_y > k_+$ the reflection of the incident electron from the lower band to the upper one is forbidden by the conservation law. But the contribution of the upper band into the wave superposition is still present in the form of the evanescent mode. The wave superposition in this case is 
\begin{eqnarray}
{\mathbf \Psi}={e^{ik_y y}\over \sqrt{2}}\left[\left( \begin{array}{c} 1\\ ie^{i\varphi_-} \end{array}\right)e^{ik_{- x} x} \right. \nonumber \\ \left.
+r_1\left( \begin{array}{c} 1\\  -ie^{-i\varphi_-} \end{array}\right)e^{-ik_{- x} x}
+g\left( \begin{array}{c} 1\\ s \end{array}\right) e^{p x} \right],  
      \end{eqnarray} 
where
\begin{eqnarray}
p=\sqrt{(k_0+\alpha)^2\sin^2\varphi_- -(k_0-\alpha)^2},~~s={k_y-p\over k_0 -\alpha}
\nonumber \\
r_1=-{s-ie^{i\varphi_-}\over s+ie^{-i\varphi_-}},~~
g=-{2i\cos \varphi_-\over s+ie^{-i\varphi_-}}.
        \end{eqnarray} 
The $z$ spin density for this wave superposition contains not only the interference contributions but also the contribution from the evanescent component $\propto e^{px}$:
 \begin{eqnarray}
s_{-z}(\vec k) =  {p\cos^2 \varphi_-\over k_0\sin\varphi_-}  [e^{2px}+\cos (2k_{-x}x)-2e^{px}\cos(k_{-x}x)]
\nonumber \\
+ {\cos\varphi_-\over k_0\sin\varphi_-}[2k_0-(k_0+\alpha)\cos^2\varphi_-]
 [\sin(2k_{-x}x)
 \nonumber \\
 -2e^{px}\sin(_{-x}kx)]. ~~
              \end{eqnarray}   
This expression is valid independently of whether the electron energy is high ($k_0>\alpha$) or low ($k_0<\alpha$).

All contributions to the $z$ spin density are odd with respect to the sign of $k_y$ and vanish in the equilibrium state. But in the presence of the voltage bias along the $y$ axis the distribution function also has  an  odd component, and spin polarization becomes possible. Let us start from the case of the ballistic regime when the voltage drop occurs at the contacts, and there is no electric field inside the sample. In the narrow interval of energies $\epsilon_F+eV>\epsilon>\epsilon_F$ around the Fermi surface only left-moving electrons with $k_y>0$ are present. They are responsible for the edge accumulation of the $z$ spin.
Bearing in mind that $eV =d\epsilon = (\hbar^2/m)k_Fdk$ the spin density $s_z(x)=s_{+z}(x)+s_{-z}(x)$  is determined by the two band contributions
 \begin{eqnarray}
 s_{\pm z}(x)={meVk_{\pm F}\over 4\pi^2 \hbar^2 k_F}\int_0^{\pi/2} s_{\pm z}(\vec k)d\varphi _\pm.
      \label{SD}       \end{eqnarray}   
Here we restrict ourselves with the limit of zero temperature, and the integration is performed over the Fermi circumferences of the two bands with the Fermi wave vectors $k_{\pm F}=|k_F \mp \alpha|$, where $k_F$ is the value of $k_0$ at the Fermi circumference.
The asymptotic behavior of the spin density is determined by the evanescent-mode contribution and at $x \to -\infty$ is given by
\begin{eqnarray}
s_z(x)={meV\over 8\pi^2 \hbar } \sqrt{\alpha\over  k_F} { 1\over k_{+F}^2k_{-F}}{1\over |x|^3}.
             \end{eqnarray} 
The total accumulated spin both for $k_F>\alpha$ and $k_F<\alpha$ is given by
\begin{eqnarray}
S_z=\int_{-\infty}^0 s_z(x)dx={meV\over 8\pi^2 \hbar  \alpha }   \left(\ln{k_F+\alpha\over |k_F-\alpha|}   -{2\alpha\over  k_F} \right). 
           \label{rho}
                 \end{eqnarray} 
This  result was obtained earlier by Zyuzin {\em et al.}\cite{Zu} for the high-energy case $k_F>\alpha$. For comparison with the collisional regime it is convenient to connect the total spin not with the voltage $V$  but with the electric current  
\begin{eqnarray}
j= {e^2n V\over \pi \hbar  }\times \left\{ \begin{array}{cc} {2k_F\over k_F^2 +\alpha^2}& \mbox{at}~k_F>\alpha\\{1\over k_F}& \mbox{at}~k_F<\alpha\end{array}\right. ,
      \end{eqnarray} 
where the 2D electron density is $n=(k_F^2+\alpha^2)/2\pi $ at $k_F>\alpha$ and $n=\alpha k_F /\pi $ at $k_F<\alpha$. Then  
\begin{eqnarray}
S_z= {mj\over 8 \pi en  } \left(\ln{k_F+\alpha\over |k_F-\alpha|}-{2\alpha\over  k_F}    \right)
\nonumber \\
\times \left\{ \begin{array}{cc} {k_F^2 +\alpha^2  \over  2k_F}& \mbox{at}~k_F>\alpha\\{k_F\over \alpha}& \mbox{at}~k_F<\alpha\end{array}\right.. 
          \label{balCur}
                 \end{eqnarray} 
In the limits of weak ($\alpha \to 0$) and strong ($\alpha \to \infty$) spin-orbit interaction this yields $S_z=(mj/24 \pi en)(\alpha^2/k_F^2)$ and $S_z=- mj/4 \pi en$ respectively. At $k_F=\alpha$ there is a logarithmic divergence, which can be cut either by the sample size or by nonlinear effects. 

Let us switch now to the collisional regime. As was explained in the Introduction, since the bulk spin current is absent, one may use the standard Boltzmann equation for a scalar distribution function $f(\vec k)= f_0(\vec k)+f'(\vec k)$, where $ f_0(\vec k)$ is the equilibrium Fermi distribution function. The stationary solution of the Boltzmann equation for the non-equilibrium distribution function $f'$ in a weak electric field  along the $y$ axis  is
\begin{eqnarray}
f'= {e\tau \vec E \over \hbar}{\partial f_0(\vec k)\over \partial \vec k} 
={eE\tau\over \hbar}{\hbar^2 k_F \over  m}\sin \varphi_\pm \delta (\epsilon-\epsilon_F),
             \end{eqnarray} 
where $\tau$ is the relaxation time for elastic scattering on defects.  The function $f'$ determines the electric current equal to $j=e^2 E\tau k_F^2/2\pi m$ and $j=e^2 E\tau \alpha k_F/2\pi m$  for $k_F>\alpha$ and $k_F<\alpha$ respectively. 
The $z$ spin densities for the two bands  instead of (\ref{SD}) are given by 
\begin{eqnarray}
 s_{\pm z}(x)={eE\tau k_{\pm F}\over 4\pi^2 \hbar }\int_{-\pi/2}^{\pi/2}\sin\varphi _\pm  s_{\pm z}(\vec k)d\varphi _\pm.
          \end{eqnarray}   
Tedious but straightforward integrations similar to those for the ballistic regime yield the total edge spin:
\begin{eqnarray}
S_z= -{mj \over 32\pi^2 e n }  {k_F^2 +\alpha^2\over k_F^4}
\left[3 (k_F^2-\alpha^2)\arctan {2\sqrt{\alpha k_F}\over k_F-\alpha}
 \right. \nonumber  \\ \left.
-{2 \sqrt{\alpha k_F}(3 k_F^2 +2k_F\alpha+3\alpha^2 )\over k_F+\alpha}+\pi (k_F^2+3\alpha^2)\right]~~
\label{colCur}
         \end{eqnarray} 
for the high-energy case $k_F>\alpha$, and 
\begin{eqnarray}
S_z = -{mj \over 16\pi^2 en  \alpha k_F}  
\left [3 (\alpha^2 -k_F^2)\arctan  {2\sqrt{\alpha k_F}\over \alpha-k_F}
   \right. \nonumber  \\ \left.
  -{\sqrt{\alpha k_F}(6\alpha^2 +6k_F^2 +4 \alpha k_F)\over \alpha +k_F}
 +4\pi \alpha k_F\right]~~
          \end{eqnarray} 
for the low-energy case $\alpha>k_F$.

\begin{figure}
\begin{center}
   \leavevmode
  \includegraphics[width=0.9\linewidth]{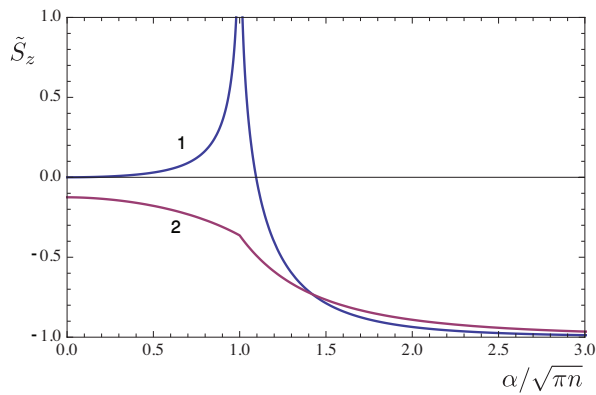} 
 \caption{(Color online) The plot of the reduced total spin $\tilde S_z=4\pi en S_z/mj$ as functions of $\alpha/\sqrt{\pi n}$. {\sl 1} -- the ballistic regime. {\sl 2} -- the collisional  regime.}
 \label{fig3}
 \end{center}
\end{figure}

When $\alpha \to \infty$ the difference between the ballistic and collisional regime vanishes. On the other hand, in contrast to the ballistic regime, in the collisional regime the accumulated spin remains finite even in the limit of zero spin-orbit coupling $\alpha \to 0$. This paradoxical result is explained by the divergence of the width $\sim 1/(k_{-x}-k_{+x})$ of the spin accumulation area in this limit. In the ballistic regime this divergence is canceled after summation over the two bands. However, our analysis is valid only if all relevant scales including $ 1/(k_{-x}-k_{+x})$ are less than the electron mean-free path. When this condition is violated the spin accumulation should go down. Figure \ref{fig3} shows the reduced total accumulated spin $\tilde S_z=4\pi en S_z/mj$ for the ballistic (curve 1) and the collisional (curve 2) regimes as functions of the density-dependent parameter $\alpha/\sqrt{\pi n}$.

Originally they connected  the spin Hall effect with bulk spin currents, and the question arises whether edge accumulation without bulk currents may be called  the spin Hall effect. A choice of terminology usually  is a matter of convention, taste, or tradition. Edge spin accumulation and spin currents require the same symmetry, and 
one may call  the edge spin accumulation without bulk currents the {\em edge} spin Hall effect. Anyway, the spin accumulation is not a method to probe spin bulk currents. A possible manifestation of the bulk spin current is an edge torque, which sometimes can be measured mechanically \cite{ES}.  It is worthwhile also  to note that in the ballistic regime the $z$-spin current vanishes not only in the bulk but everywhere the accumulation area including. So the edge spin Hall effect is not stipulated by 
 violation of the spin conservation law since the latter can be not violated.

In order to compare the edge and the bulk spin Hall effects, we scale the latter using the ``universal'' spin conductivity $\sigma_{SH} = j^z/E=e/8\pi $, though in reality this is far from being universal \cite{Engel}. Here $j^z$ is the bulk  current  of the $z$ spin. Assuming that at the edge the bulk spin current is fully compensated with spin diffusion current  \cite{Dy08,DP}, the total accumulated spin is $j_z \tau_s=eE\tau_s/8\pi$, where $\tau_s$ is the spin relaxation time. So ratio of the edge to the bulk spin  Hall effect  is $\sim \tau /\tau_s$.  

For comparison with the spin Hall effect observed in the 2D hole gas \cite{Wund,Nomura} on may use  $\tau \sim 10\hbar /E_F=20/k_F v_F$, $n= 2\times 10^{12}$ cm$^{-1}$, and  the accumulation area width 10 nm given by Nomura {\em et al.} \cite{Nomura}. Then  the total spin accumulated  due to the edge spin Hall effect  at $\alpha \to 0$ is about 70 \% of the experimental value. Therefore, the interpretation of this experiment in the terms of the bulk spin currents probably must be reconsidered. 

In summary, in a system with spin-orbit interaction an electric field can lead to spin accumulation at sample edges normal to the field even without bulk spin currents (edge spin Hall effect). It has been demonstrated for a 2D electron gas in the collisional regime. Therefore observation of edge spin accumulation  cannot be a probe of bulk spin currents, and other methods must be used for their detection \cite{spin}. 

The work was supported by the grant of the Israel Academy of Sciences and Humanities.


\begin{thebibliography}{99}
\bibitem{Engel}  H.-A.  Engel, E.~I.~Rashba, and B.~I.~Halperin,
 in {\sl Handbook of Magnetism and Advanced Magnetic Materials}, edited by
 H.~Kronm\"uller and S.~Parkin (Wiley, New York, 2007), Vol.~5, pp.  28--58;    \eprint{cond-mat/0603306}.

  \bibitem{Dy08} M.~I.  D'yakonov  and A.~V.~Khaetskii,   in {\sl Spin Physics in Semiconductors}, edited by M.~I.  D'yakonov (Springer, Berlin, 2008),  pp.211--243.
  
  \bibitem{DP} M.~I.  D'yakonov  and V.~I. Perel,
Pis'ma Zh. Eksp. Teor. Fiz.  {\bf 13}, 657 (1971) [JETP Lett.  {\bf 13}, 467 (1971)].

  \bibitem{spin} E. B. Sonin,     arXiv:cond-mat/0807.2524.
  
  \bibitem{R} E. I. Rashba,  Phys. Rev. B {\bf 68}, 241315(R) (2003).

\bibitem{ES} E. B. Sonin,  Phys. Rev. Lett.  {\bf 99}, 266602 (2007).

\bibitem{Nik} B. K. Nikoli\'c, S. Souma,  L. P. Z\^arbo, and J. Sinova, Phys. Rev. Lett. {\bf 95}, 046601  (2005).

\bibitem{Usaj} G. Usaj and C. A. Balseiro, Europhys. Lett., {\bf 72},  631 (2005)

\bibitem{Zu} V. A. Zyuzin, P. G. Silvestrov, and E. G. Mishchenko, Phys. Rev. Lett. {\bf 99}, 106601  (2007).

\bibitem{Kato}  Y. K. Kato,  R. C. Myers, A. C. Gossard, and D. D. Awschalom, Science {\bf 306}, 1910 ( 2004).

\bibitem{Wund} J. Wunderlich,  B. Kaestner, J. Sinova, and T. Jungwirth,  Phys. Rev. Lett. {\bf 94}, 047204 (2005). 

\bibitem{Nomura} K. Nomura, J. Wunderlich,  J. Sinova,  B. Kaestner, A.~H.~MacDonald, and T.~Jungwirth,  Phys. Rev.B {\bf 72}, 245330 (2005). 

\bibitem{DK} M.~I.  D'yakonov  and A.~V.~Khaetskii,  Sov. Phys. JETP   {\bf 59}, 1072 (1984).
\bibitem{Shyt} A. V. Shytov, E. G. Mishchenko, H.-A. Engel, and B. I. Halperin,  Phys. Rev. B {\bf 73}, 075316 (2006).
\end{thebibliography}
\end{document}